# Hydrogel microwells with light-controlled reversible closure

Qifei Ma[1], David Urban[2,3,4], Stefano Gabetti[5], Beatrice Masante[5], Huaizhou Jin[6], Federica Galvagno[7], Diana Massai[5], Alberto Puliafito[7], Shangzhong Jin[1*], Denis Garoli[1,8,9*], Emiliano Descrovi[4*]

[1]College of Optical and Electronic Technology, China Jiliang University; Hangzhou 310018, China.

[2]Department of Electronic Systems, Norwegian University of Science and Technology; Trondheim, NO-7491, Norway.

[3]SINTEF Digital, P.O. Box 124 Blindern, Oslo, Norway

[4]Department of Applied Science and Technology, Politecnico di Torino; Torino, 10129, Italy.

[5]Department Mechanical and Aerospace Engineering, Politecnico di Torino; Torino, 10129, Italy.

[6]College of Physics, Zhejiang University of Technology; Hangzhou, 310027, China

[7]Department of Oncology, University of Torino; Orbassano, 10126, Italy

[8]Istituto Italiano di Tecnologia; Genova, 16163, Italy

[9]Dipartimento di Scienze e Metodi dell'Ingegneria, Università degli Studi di Modena e Reggio Emilia; Reggio Emilia, 42122, Italy.

*Corresponding author. Email: denis.garoli@unimore.it (D.G.); jinsz@cjlu.edu.cn (S.J.); emiliano.descrovi@polito.it (E.D.)

**Abstract.** We present a light-responsive hydrogel nanocomposite engineered into arrays of micrometer-scale wells that can be selectively and sequentially closed and re-opened via laser illumination. Polarization-controlled light exposure induces anisotropic surface deformations, leading to the formation of protrusive flaps sealing the wells. Owing to the intrinsic elasticity and anti-adhesive properties of the hydrogel matrix, the deformation process is partially reversible, allowing flap retraction and restoration of the original well geometry. This platform facilitates contactless, on-demand trapping and release of microscale objects using a standard optical microscopy configuration. As a proof of concept, we demonstrate the controlled manipulation of a single polystyrene microbead using optical tweezers, including bead positioning within a well, light-triggered closure, and subsequent reopening to release the particle into the surrounding aqueous environment.

## Introduction

Photo-deformable hydrogels constitute a versatile class of adaptive soft materials in which optical inputs are transduced into mechanical motion through a variety of physicochemical mechanisms.[1,2] Depending on network architecture and molecular organization, illumination can induce deformation via photothermal,[3,4] or photochemical processes[5,6] as well as through photoinduced molecular reorientation,[7] each characterized by distinct response times, reversibility, and symmetry of actuation. Photothermal mechanisms are among the most widely employed, relying on embedded light-absorbing inclusions—such as gold nanoparticles[8] or rare-earth oxide particles[9] —to generate localized heating and trigger a volume phase transition in thermoresponsive polymers (e.g., PNIPAm gels[10]). In structurally homogeneous networks, this typically produces isotropic deformation. However, the presence of mechanical constraints,[11] tailored geometries,[12] or compositional gradients[13,14] can yield more complex, anisotropic responses.

A fundamentally different actuation regime arises from photoinduced molecular reorientation and anisotropic stress generation, particularly in azobenzene-containing systems[15] and liquid-crystalline polymer networks.[16,17] In these materials, the mechanical response is inherently vectorial: both the magnitude and direction of strain depend on the polarization of the incident light relative to the molecular alignment.[18,19] Because deterministic control of actuation symmetry and direction requires predefined molecular alignment, the deformation mode is effectively inscribed during fabrication and cannot be altered afterward.

To address this constraint, a recent study by some of the present authors introduced an amorphous photoresponsive composite capable of directional actuation governed not only by intensity but also through the polarization state of the incident light.[20] This concept builds on the work of Ryabchun and Bobrovsky,[21] who demonstrated that microdomains of a liquid-crystalline PAAzo copolymer dispersed within a passive elastomer matrix exhibit pronounced, reversible directional deformation. Extending this approach, it was shown that azopolymer micro- or nanoparticles embedded at sufficiently high concentrations can effectively transmit their intrinsic light-driven deformation to the surrounding elastomeric host. The collective effect produces macroscopic, polarization-dependent actuation that is both reversible and directionally controlled, even in the absence of long-range molecular order in the composite.

In an attempt to demonstrate the potential of this approach in unconventional tasks, we address here the actuation of micro-well arrays made of a PVP-based hydrogel containing azopolymer nanoparticles (hereafter termed azoGel).[22] We demonstrate that wells with lateral dimensions up to 20 μm and depths of 6–8 μm immersed in water, can be reversibly closed and reopened using different forms of polarized light. While well sidewalls remain largely preserved during actuation; closure occurs via the emergence of protrusive flaps from the upper rim that extend over — and seal — the aperture. Said flaps can be contracted to re-open the well, by employing illumination with an orthogonal polarization state. Two illumination schemes are investigated and compared: uniformly linearly polarized beams and doughnut-shaped beams with either radial or azimuthal polarization. As a proof of concept for an all-optical micromanipulation, a single polystyrene bead is optically trapped, inserted into a well, segregated within, and subsequently released into the surrounding medium upon well re-opening.

## Results and Discussion

Square microwells were fabricated following the protocol described in the Materials and Methods section. The materials used are known to exhibit reversible photo-actuation governed by the polarization of incident light.[22] In an initial experiment, individual microwells are illuminated with a linearly polarized beam in an inverted microscope setup (see Materials and Methods). Figure 1(a) presents a time-lapse sequence acquired during irradiation. When the polarization is oriented vertically relative to the camera frame (y-axis), two opposing flaps emerge along the polarization direction, extending over the underlying water-filled cavity. While polarization-driven anisotropic deformations in azopolymer-based systems are well documented, here we observe supra-micrometric protrusions of the azo-hydrogel that extend well beyond the bulk material boundary.  With continued illumination, the flaps progressively elongate until they meet, effectively sealing the well. Rotating the polarization by 90° induces flap contraction, resulting in partial reopening of the microwell. Although the aperture is restored toward its initial state, repeated cycling between vertical and horizontal polarization drives the system toward a progressively more closed configuration, reducing the effective opening to approximately 17% of its original area after two cycles (Supplementary Figure S1). Well open area was measured by digitally segmenting brightfield images with Ilastik.[23] In Supplementary Movie M1, the whole sequence is provided.

To further characterize the deformation dynamics, we computed the velocity field of the material (see Materials and Methods section). Representative frames during closing and reopening are shown in Fig. 1(b–e), where velocity vectors are overlaid on optical images alongside the in-plane angular distribution of the velocity direction, defined as $\theta_v = \mathrm{tg}^{-1}\left(\frac{v_y}{v_x}\right)$. During closure (Figs. 1b,d), velocity vectors associated with flap motion align predominantly with the polarization axis (y-direction) and point inward toward the well center. This coordinated motion is reflected in the $\theta_v(x, y)$ distribution, which displays a sharp sign change from $\theta_v = \frac{\pi}{2}$ to $\theta_v = -\frac{\pi}{2}$ across a horizontal line through the well center. Upon polarization rotation (Figs. 1c,e), the vectors remain largely oriented along y, but the $\theta_v$ distribution inverts vertically. Reopening arises from flap contraction along y-direction accompanied by a later elongation along x-direction. This behavior is attributed to elastic restoring forces within the matrix, enabling sequential orthogonal deformations to remain partially reversible and non-cumulative,[20] in addition to the anti-sticking properties of the hydrogel.

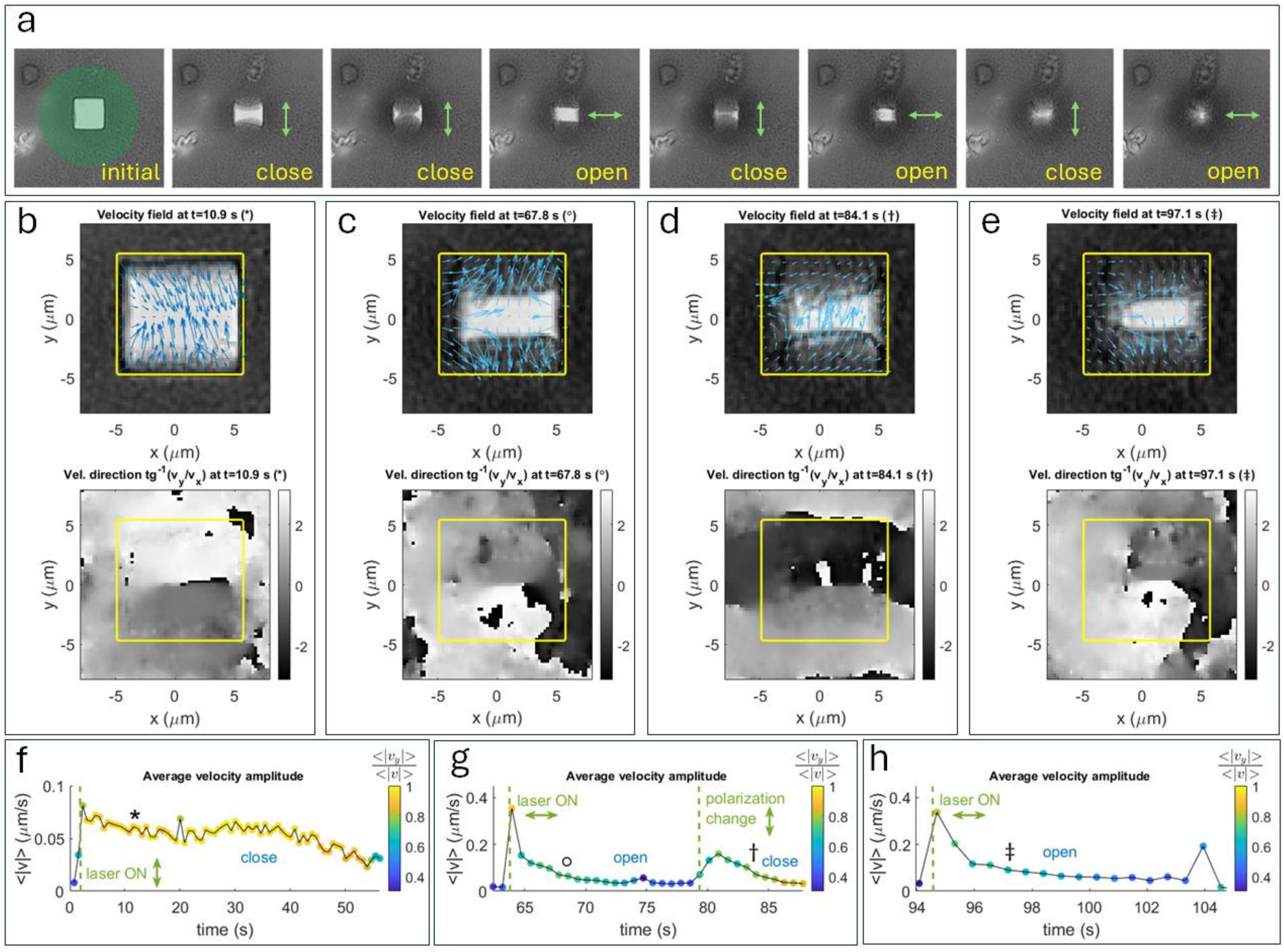

**Figure 1. Photo-actuation with linear polarization.** (a) Sequence of frames showing a hydrogel well (10 μm side) continuously illuminated with alternating linear polarization states with orthogonal orientations, as indicated; (b-e) velocity fields (arrow plots) superposed to raw optical images at illustrative time points during closing/opening of the well and corresponding distributions of the velocity direction with respect to the horizontal (x-axis), calculated as $\theta_v = \mathrm{tg}^{-1}\left(\frac{v_y}{v_x}\right)$. Well boundaries are outlined by the yellow square; (f-h) average velocity amplitude calculated over time, across the whole well area. Marker color is proportional to the ratio of the average velocity y-component to the average velocity $\frac{\langle |v_y| \rangle}{\langle |v| \rangle}$. Laser intensity $I \cong 10$ W/cm$^2$. See Supplementary Movie M1.

It is worth observing that reopening would be hampered if flap contraction occurred at the same rate as the formation of new protrusions from the orthogonal well edges. Stated otherwise, if flap contraction along the y-direction would occur at the same speed as flap extension along the x-direction, the well aperture could not be recovered in size, since what would be gained in the y-direction would be lost in the x-direction. To further elucidate the mechanisms governing the closure and reopening of wells, we analysed the time evolution of the spatially averaged velocity magnitude over the entire well area. The first closing stage occurs over approximately 60 s, with mean velocities ranging from 30 $\mathrm{nm/s}$ and 82 $\mathrm{nm/s}$, substantially dominated by the $\langle|v_y|\rangle$ component (Fig. 1f). Upon rotating the polarization to the x-direction to initiate reopening, the average velocity briefly rises to values as high as 350 $\mathrm{nm}/s$ over a few seconds, while remaining primarily oriented along the y-direction (Fig. 1g). During this interval, rapid flap retraction produces substantial reopening of the well. Average velocity subsequently decreases to about 35 $\mathrm{nnm/s}$; in this regime, the contribution from $v_y$ progressively diminishes, likely reflecting the emergence of a $v_x$ component associated with the early formation of a new pair of flaps aligned along the x-direction. When the polarization is rotated (well closing), the velocity increases abruptly, the y-oriented flaps rapidly protrude again, and the well recloses within approximately 15 s—representing a significant acceleration compared to the initial closing stage. A similar behaviour is observed during the second reopening cycle (Fig. 1h), characterized by fast contraction of the y-oriented flaps, with the average velocity initially aligned along y-direction and gradually becoming more x-oriented. The complete sequence corresponding to Fig. 1 is shown in Supplementary Movie M1. An additional illustrative case featuring well closing and opening is provided in Supplementary Figure S2 and Supplementary Movie M2, showing a similar behaviour. In both examples, the flap shapes demonstrate a deformation that is more pronounced in the central part, as opposed to the lateral sides. We speculate that this might be due to the emergence of shear forces at the well corners in the early phase of flap formation, when the azo-hydrogel starts to protrude from the bulk. This mechanism, together with the non-sticking characteristics of the material (not observed in conventional pure azopolymers[24,25]) can explain the velocity difference observed between the first closing step and the subsequent re-opening and re-closing steps (Figs. 1f-h), when shear forces in flap protrusion and contraction are likely to be weaker. Although well's closure and reopening are enabled by such velocity asymmetry between the initial and subsequent hydrogel deformations, repeated cycling ultimately results in near-complete and irreversible sealing after approximately three to four cycles, accompanied by a progressive reduction of the effective aperture (Supplementary Figure S1). This drawback can be partially alleviated by operating on the illumination pattern used to promote the photo-actuation. Indeed, illumination with uniformly polarized fields is inherently suboptimal, as it drives actuation along only two orthogonal directions, whereas efficient and fully reversible well operation would likely require radially symmetric flap actuation.

To address this aspect, illumination schemes with spatially varying polarization should be explored, particularly axis-symmetric configurations such as radial and azimuthal polarization. These polarization states are well established in laser optics and can be generated through several approaches, including the use of holographic or diffractive plates,[26,27] Spatial Light Modulators[28] or Liquid-Crystal elements,[29] as in the present case (see Materials and Methods section). Figures 2a–c show the hydrogel surface after a few seconds of exposure to a weakly focused laser beam with linear, azimuthal, or radial polarization, respectively. The images are extracted from the real-time sequence provided in Supplementary Movie M3. Consistent with observations in other azopolymeric systems, this hydrogel exhibits surface patterning governed by the local orientation of the laser electric field, producing wrinkles aligned perpendicularly the local polarization direction.[30] Notably, when the polarization is configured as a superposition of radial and azimuthal states, wrinkles organize into spiral-like structures (Fig. 2d), which may promote vortex-like mass transport,[31] as illustrated in Supplementary Movie M4.

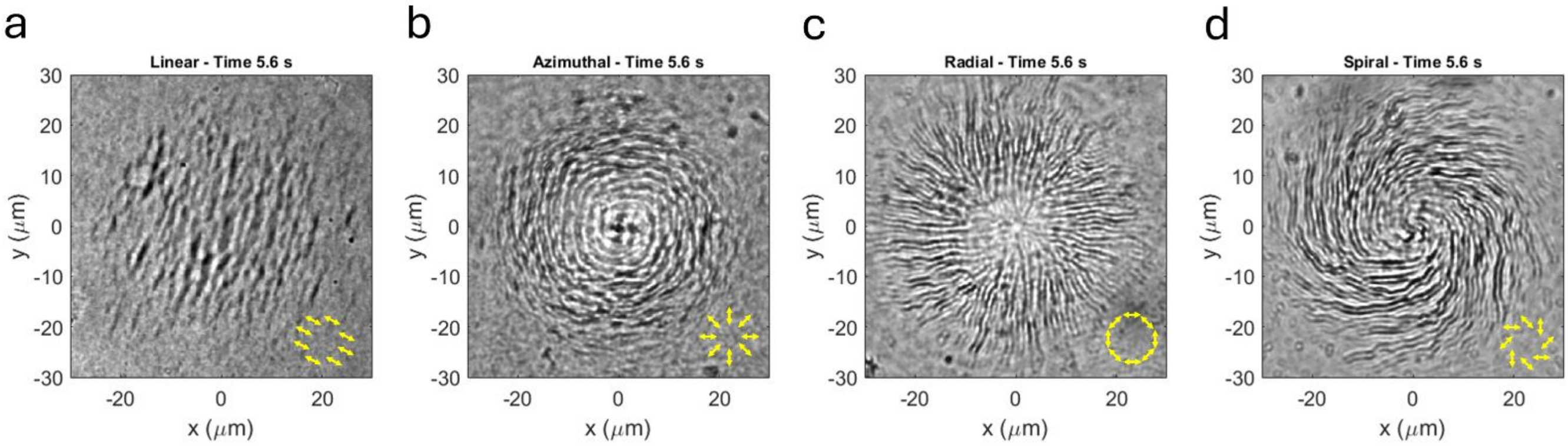

**Figure 2. Hydrogel patterning with several polarization distributions.** Optical images of the hydrogel surface after being irradiated for few seconds with a weakly-focused beam having linearly-polarization (a), radial polarization (b), azimuthal polarization (c) and a mixed polarization state given by superposition of azimuthal and radial, leading to spiral-oriented wrinkles. Images taken from Supplementary Movie M3.

Figure 3a illustrates a representative closing–opening cycle achieved using a weakly focused doughnut-shaped beam with an axis-symmetric polarization profile. The transverse beam size is chosen such that the central dark region aligns with the well aperture, while the surrounding bright ring illuminates the well sidewalls (Supplementary Figure S3). Under radially polarized illumination, the hydrogel deforms outward along radial directions, producing approximately four flaps that extend toward the center of the well. Reopening is then induced by switching to the orthogonal polarization state, implemented as an azimuthally polarized doughnut beam. In this configuration, the flaps undergo radial contraction, resulting in near-complete restoration of the well opening. Worth of noting, circularly polarized excitation would promote well closure but would not enable reopening.[32,33] Figs. 3b,d present representative frames acquired at selected time points during the closing phase. In this regime, the velocity field is predominantly radial, with vectors directed toward the well centre, corresponding to a mainly negative radial velocity component $v_r$ in the regions associated with flap deformation. Conversely, in Figs. 3c,e flaps expand outward, yielding clearly positive values of $v_r$. Repeated recovery of the well aperture is substantially improved compared with the case of homogeneous linear polarization, leading to the clear aperture of the well to be recovered up to 70% after three cycles (Supplementary Figure S1).

The spatially averaged velocity magnitude over the entire well area (Fig. 3f) exhibits a trend similar to that observed in the previous case, with higher velocities during flap contraction after the very first expansion. For instance, the first reopening event beginning at $t = 30$ s occurs at velocities between 100 and 200 $\mathrm{nm/s}$, which are sustained for a longer duration than during the preceding closing step, despite constant illumination intensity. With successive expansion–contraction cycles, the peak actuation velocity progressively increases, and elevated velocities persist for longer intervals. We attribute this enhanced mobility to a progressive thinning of the hydrogel regions involved in actuation, relative to the initial illumination stage, during which the flaps need to be first formed (against shear forces at the well corners) and then protruded from the bulk material surrounding the well aperture.

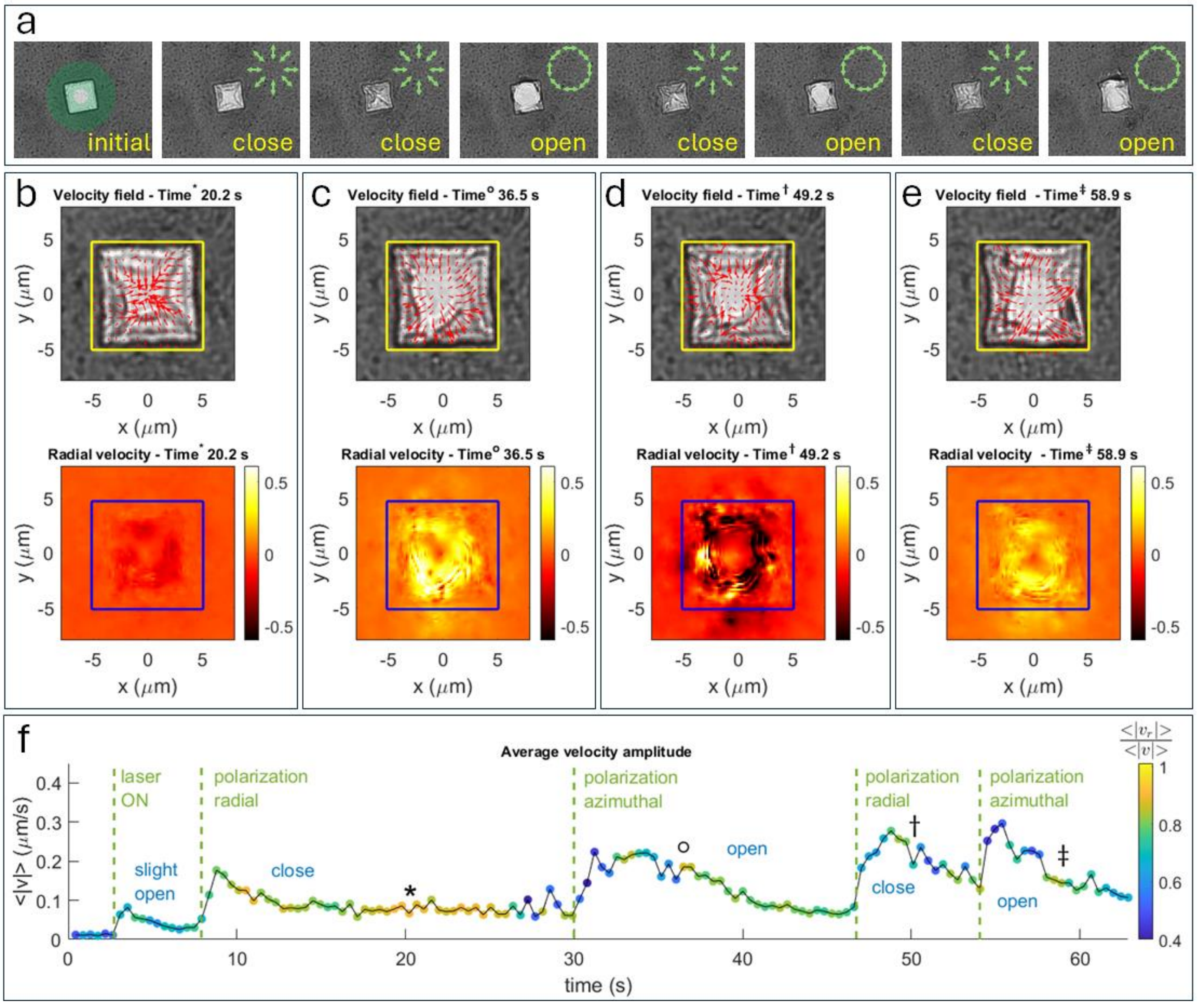

**Figure 3. Photo-actuation with axis-symmetric polarization.** (a) Sequence of frames showing a hydrogel well (10 μm side) continuously illuminated with alternating radial and azimuthal polarization states, as indicated; (b-e) velocity fields (arrow plots) superposed to raw optical images at illustrative time points during closing/opening of the well and corresponding distributions of the radial component of the velocity, calculated as $v_r = v_x \cos\varphi + v_y \sin\varphi$, with $\varphi$ being the polar angle with respect to the x-axis; (f) average velocity amplitude calculated over time, across the whole well area. Marker color is proportional to the ratio of the average velocity radial component to the average velocity $\frac{\langle|v_r|\rangle}{\langle|v|\rangle}$. Laser intensity $I \cong 10$ W/cm$^2$. See Supplementary Movie M5.

Flap protrusion over the well aperture is clearly evidenced by pseudo–bright-field transmission images and z-stack fluorescence microscopy acquired using a confocal setup (Fig. 4; see Materials and Methods). In pseudo–bright-field mode, the pristine well  well appears brighter due to less absorption by azopolymer particles compared with the surrounding thick gel layer (Fig. 4a). Cross-sectional fluorescence images reveal strong emission from the azo-gel at the top surface, with weaker signal originating from the hydrogel bulk, because of absorption, as mentioned. A thin residual azo-gel layer at the well bottom, in contact with the glass substrate, provides a weak fluorescence signal enabling the estimation of the well depth at approximately 6.5 µm, the sample-to-sample variations being up to 1.5 µm.

Upon illumination with linearly polarized light, flaps extend several micrometers without collapsing (Fig. 4c), and cross-sectional views show these structures suspended above the well cavity (Fig. 4d). More effective isolation of the well volume from the surrounding aqueous environment is achieved using radially-polarized illumination, which promotes tighter closure of the well lid (Figs. 4e,f). In this configuration, fluorescence from the well bottom is strongly attenuated due to absorption by the closed lid, preventing a clear visualization of the cavity base.

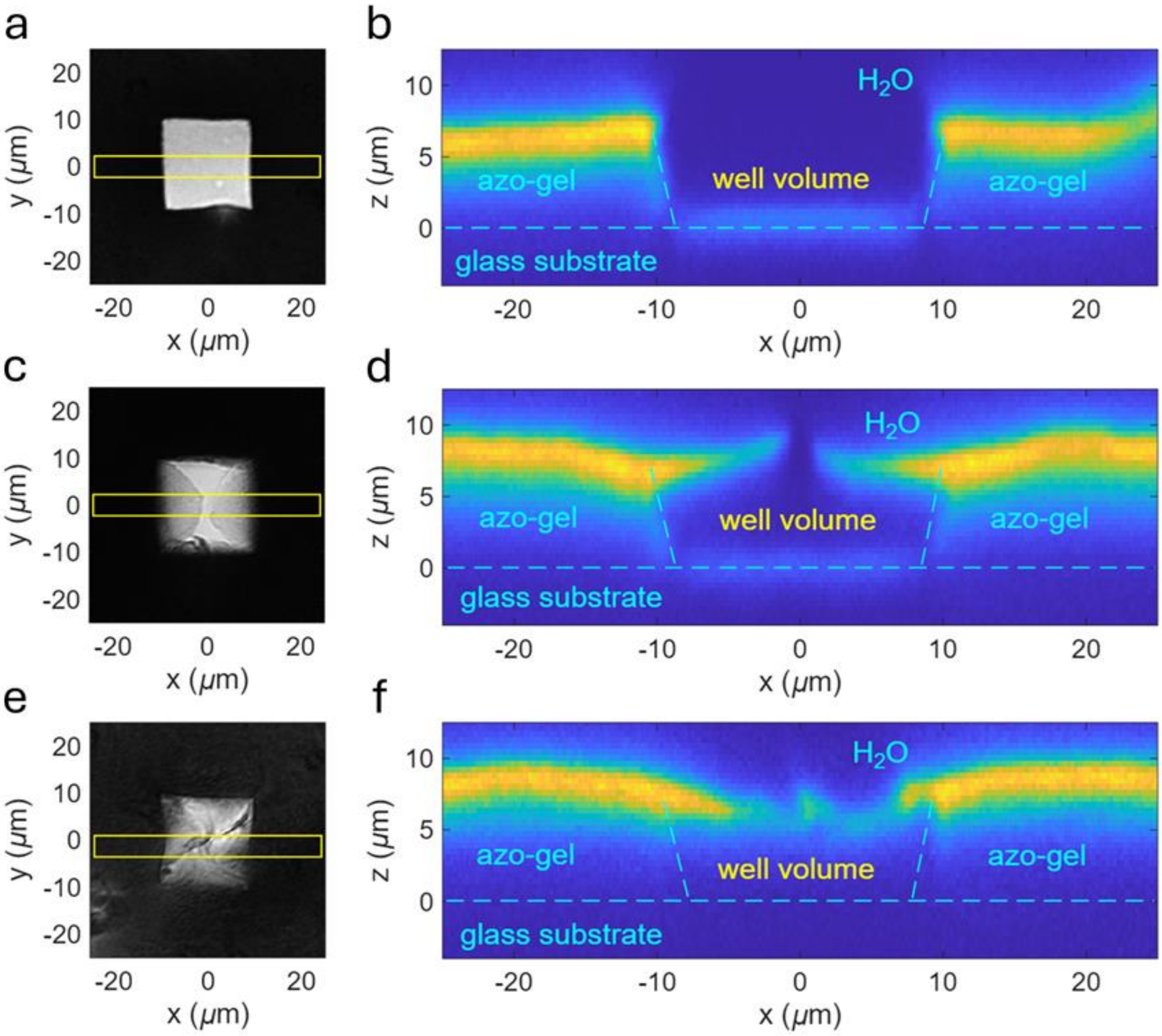

**Fig. 4. Cross sections of open and closed wells.** Pseudo bright-field top views and confocal fluorescence cross-sectional views of a pristine well (a, b), a partially closed well illuminated with linearly polarized beam (c, d), a completely closed well illuminated with radially polarized beam (e, f). Cross-sectional views are obtained upon

integration of z-stacks over the yellow rectangle shown in the pseudo bright-field images. Well side length: 20 μm.

Having demonstrated reversible closure and reopening of wells with lateral dimensions up to 20 μm, we next present a proof-of-concept application in which a 1.7 μm–diameter polystyrene microbead dispersed in water is confined within a microwell and subsequently released using optical tweezers (OT). To facilitate integration within the OT setup, well actuation was performed using a linearly polarized illumination beam, at the cost of a less effective sealing. Moreover, for a reliable bead segregation and release, wells were functionalized with Bovine Serum Albumin (see Materials and Methods) in order to avoid beads to adhere to the well internal surfaces.

Representative frames of the full procedure are shown in Fig. 5a. First, a single polystyrene bead is trapped by the OT and guided into an open well. The well is then actuated using linearly polarized illumination along the x-direction to initiate closure, which proceeds in two successive steps. Although the well reaches a partially closed state, the bead remains clearly visible beneath the protruding flap. Laser is then switched on again, with a rotated polarization by 90° to induce reopening of the well. Following this step, the bead is again accessible from the outside and can be released in the water medium using the OT.

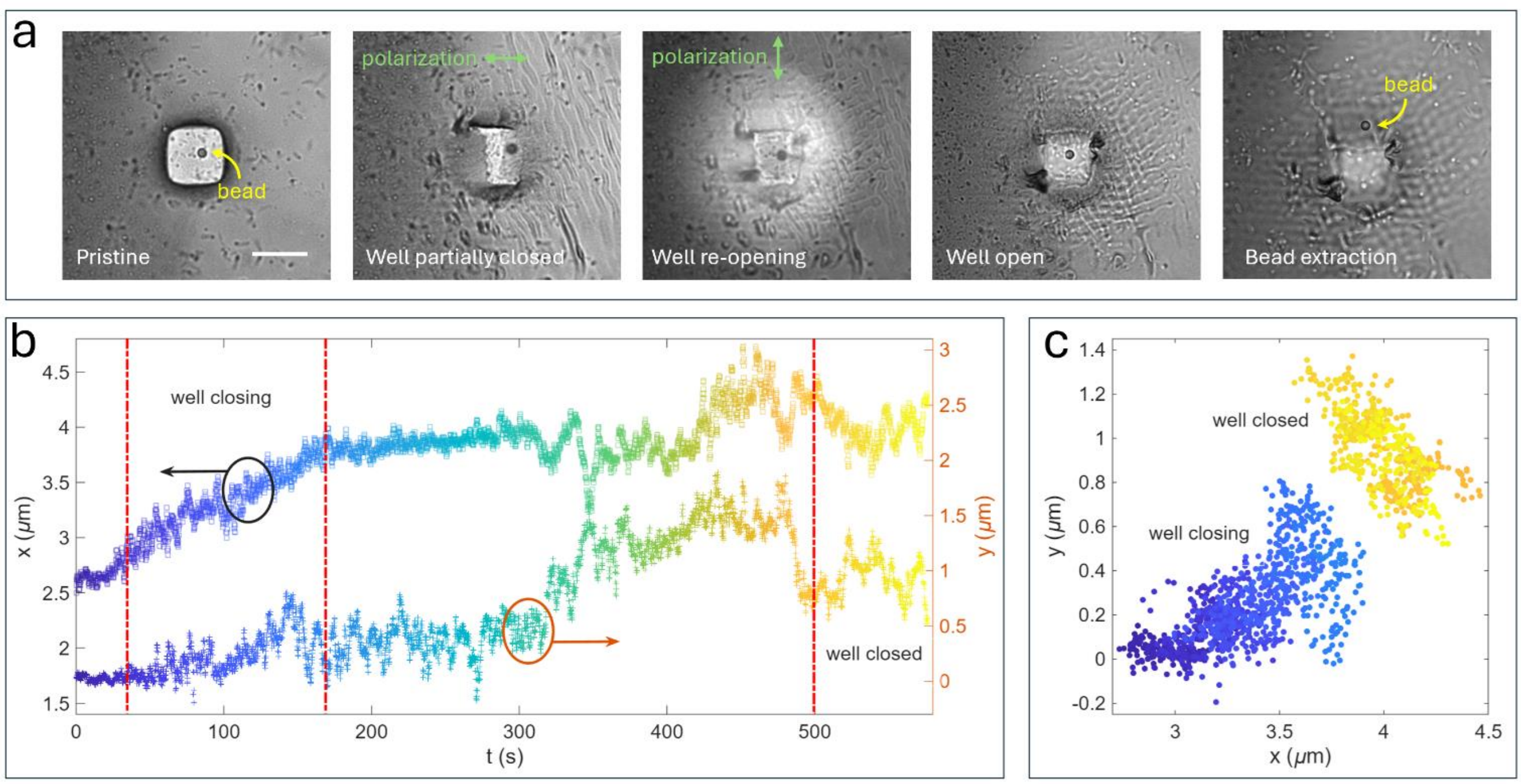

**Fig. 5. Polystyrene bead manipulation.** (a) Frame sequence for the all-optical manipulation of a polystyrene

bead. From left to right: insertion of the bead within an open well, closing of the well via linearly-polarized beam illumination, re-opening of the well via (orthogonal) linearly polarized beam; extraction of the bead from the well. Scale bar is 10 μm; (b) x-position (squared markers) and y-position (cross markers) of the bead center of mass tracked over time during the well closing; (c) scatter plots of the bead xy-position during the first illumination step of well closing and after the well is closed. Well side length: 10 μm.

A natural question is whether the bead becomes mechanically constrained within the well or retains some freedom of motion in the cavity volume. To address this, we tracked the bead position over time by extracting the xy-coordinates of its center of mass. The bead exhibits Brownian motion, wandering across a region of approximately 2 μm in both lateral directions (Fig. 5b). This behavior indicates minimal adhesion to the hydrogel, which is consistent with effective BSA functionalization. More importantly, comparison of position fluctuations during and after laser-induced closure of the well reveals no significant differences in the distribution spread (Fig. 5c), indicating that the bead is free to move within the well cavity, with no strong mechanical constrains played by either the well sidewalls or the protruding flaps, when segregated within.

**Conclusion**

In conclusion, we have introduced a light-responsive, azo-doped hydrogel platform capable of controlled and partially reversible closure and reopening of micrometer-scale wells under aqueous conditions. Polarized illumination drives the formation of protrusive flaps oriented along the local polarization direction, enabling spatially defined actuation. Switching between orthogonal polarization states induces elastic retraction and partial reopening, restoring up to 75% of the original clear aperture. The approach is entirely contactless and readily implemented within standard optical microscopy configurations, as demonstrated using a commercial optical tweezer system to achieve all-optical trapping, confinement, and subsequent release of a single micro-object. Because the actuation symmetry is governed by light polarization rather than pre-encoded structural anisotropy, the platform offers a high degree of user-defined control, thereby overcoming key limitations associated with alignment-dependent liquid-crystalline materials.

Beyond proof-of-concept validation, this system presents promising opportunities in light-regulated microfluidics. Recent progress in hydrogel-based photoresponsive valves for flow control[34] has underscored the importance of remotely addressable, low-power, and reversible gating strategies for lab-on-chip technologies. In this context, the microwell architecture described here can function as an optically actuated microvalve, enabling localized and programmable capture or release of micro-objects within microchannels or microchambers, and conceptually extendable to adaptive membrane systems.[35] Such capabilities support dynamic compartmentalization, on-demand mixing, and spatiotemporal regulation of microscale transport in aqueous environments. In biomedical applications, light-responsive organic materials have been explored as 2D substrates[36] and 3D environments[37] for cell culture. Applied to microwell platforms,[38] the reversible sealing of confined volumes further suggests applications in studies focused on growth, proliferation and metabolism control of cell aggregates[39,40] or single-cells.[41] Future integration with microfluidic circuitry and optical manipulation tools could enable programmable arrays of light-driven gates for parallelized and reconfigurable microsystems.

## Materials and Methods

**Azo-particle preparation**. An azopolymer solution was prepared by dissolving poly(Disperse Red 1 methacrylate-co-methyl methacrylate) (p(DR1m-co-MMA), ~15 mol% dye monomer; Sigma-Aldrich) in chloroform at a concentration of 10 mg mL$^{-1}$, subsequently combined with an aqueous 0.5 wt% poly(vinyl alcohol) solution (PVA; Sigma-Aldrich, MW 13–23 kDa, 87–89% hydrolysed) at a volume ratio of 1:12.5. Emulsion is then obtained in a sealed container using an ultrasonic bath (Elmasonic P30H, 37 kHz, 100%) for several hours at 45–55 °C, with intermittent manual agitation. Particle formation was induced by pouring the emulsion into a beaker and gradually raising the bath temperature to 80 °C to force chloroform evaporation. Excess PVA was removed by centrifugation, followed by resuspension of the particles in deionized water; this purification step was repeated five times. Finally, the particle suspension was concentrated by adjusting the volume of deionized water during the final resuspension, yielding an azopolymer particle dispersion with an approximate solids content of 30 wt%.

**Structure fabrication.** The particle precursor dispersion (30 wt% azopolymer particles stabilized with poly(vinyl alcohol), PVA) was combined with an aqueous solution of poly(vinyl pyrrolidone) (PVP; 10 wt%, MW 360 kDa; Sigma-Aldrich). The mixing ratio was adjusted to achieve the targeted nanoparticle-to-PVP solids composition (NP:50% PVP). The resulting mixtures were degassed under vacuum and cast between a cleaned glass microscope slide and a polydimethylsiloxane master mold (PDMS; Sylgard 184, Dow Corning). The PDMS molds contained inverse microstructures corresponding to negative replicas of well arrays fabricated by standard SU-8 photolithography. Following solvent evaporation, the PDMS mold was removed, and the microstructured films were annealed on a hot plate at 135 °C for 10 min. under dry conditions, exceeding the glass transition temperature of p(DR1m-co-MMA) ($T_g \approx 105$ °C). Prior to use, samples were immersed in deionized water for 1 h.

**Experimental setups**. An inverted microscope (Nikon Ti2) is equipped with an additional light path to allow a laser beam to be directed toward the sample through the imaging objective. On the filter wheel, a dichroic mirror (Thorlabs, DMLP550R, 550 nm cut-off) combined with a long-pass spectral filter (Thorlabs, FELH0550) are employed to facilitate the laser irradiation of the sample and reduce reflection onto the imaging camera. Laser light is emitted from a CW doubled-frequency Nd:YAG laser emitting at 532 nm (Torus 532, Novanta Precision Manufacturing), then injected into a single-mode fiber and expanded (beam diameter from 5 mm to 20 mm). Polarization is controlled by the use of a linear polarizer and a half-waveplate. When radial/azimuthal polarization is employed, an additional Liquid Crystal element (Zero-Order Vortex Half-Wave Plate, Thorlabs WPV10L-532) is inserted in the optical path. By varying the divergence of the expanded beam, the size of the illumination spot can be controlled. Typically, illumination areas are limited to a maximum diameter of up to 30 μm, with intensity of about 10 $W/cm^2$.

**Confocal images**. Confocal and pseudo bright-field images are taken with a Leica Stellaris 5 (Leica Microsystems GmbH, Wetzlar, Germany) inverted microscope. Azo-gel wells fabricated on a glass slide are put face-down within a Petri dish filled with deionized water. A pair of 100 μm-thick tape stripes are inserted as spacers between the Petri dish bottom and the sample glass substrate, to avoid the azo-gel structure directly pressed against the dish. Illumination is provided by raster scanning a focused laser spot (wavelength $\lambda_{ex}$ =570 nm) through the sample, using a water immersion 25X objective with NA=0.95 (HC FLUOTAR L VISIR). Fluorescence is collected by the same objective at wavelength 580 nm $< \lambda_{em} <$ 710 nm. Pseudo bright-field images are formed by collecting transmitted excitation light by means of the microscope condenser lens equipped with a dedicated photomultiplier.

**Calculation of the velocity fields.** Velocity fields are calculated by processing image sequences by means of the Optical Flow technique implemented in MATLAB R2025b. More specifically, the Horn-Schunck method is employed, with a "smoothness" parameter between 0.02 and 0.05. Raw images were recorded at 5 fps, but Supplementary Movies M1-3 are played at 2.5 frames per second.

**Force measurements**. Measurements have been performed with the Optical Tweezer Sensocell setup mounted on a Nikon Ti2 microscope (objective: Nikon CFI Plan Apo VC 60XC WI with magnification 60x and NA=1.2). In contrast to conventional force spectroscopy techniques that rely on indirect calibration, this optical tweezers platform measures forces by directly detecting changes in light momentum resulting from the interaction with an arbitrarily-shaped trapped particle.[42] This approach eliminates the need for in situ trap stiffness calibration, as the force is derived from first principles of momentum conservation. To prepare for such measurements, the system requires a stable laser source, which is achieved by turning on the laser at considerably high power at least 30 minutes before the experiment and controlling the final trap power with a rotating half-wave plate. Accurate force detection further depends on the precise alignment of the optical path. This is accomplished using an empty microchamber for calibration: a water droplet is first placed on the water-immersion objective, and a droplet of immersion oil is applied on top of the upper glass slide covering the sample. The collecting lens of the force sensor unit is then carefully lowered until it makes contact with the oil droplet, forming a continuous, unobstructed optical path. It is critical to ensure there are no air bubbles within the oil droplet, as these can distort the light and directly compromise the fidelity of the force measurements. Polystyrene beads (1.74 μm diameter) were first injected into an open microfluidic chamber made of two glass slides separated by 200 μm spacers. The bottom slide is coated with a uniform layer of azo-doped hydrogel. After 20-minutes, beads in adhesion with the hydrogel layer were optically trapped, with the trapping laser power gradually increasing until the adhered particle was detached. At this point, the applied optical force components $F_x$ and $F_y$ were recorded and the total adhesion force $F = \sqrt{F_x^2 + F_y^2}$ was calculated. Same protocol was employed to measure adhesion forces between individual DR1M particles suspended in water and the azo-doped hydrogel. Results are shown in Supplementary Figure S4. The surface of the azo-doped hydrogel was then modified through interface passivation using Bovine Serum Albumin (BSA).[43] Following surface modification, after 20 minutes of static settlement, most polystyrene beads remained suspended in solution, with negligeable sticking. Adhesion forces after BSA functionalization were too low to be reliably measured with our setup.

**Data Availability**

The data and analysis codes that support the findings of this study can be made fully available upon reasonable request.

**Competing interests**

D.U. and E.D. are inventors on a patent application (Italian Patent, Application No: 102025000006681) filed by Politecnico di Torino and NTNU Technology Transfer AS on March 31, 2025. The authors declare that they have no other competing interests.

**Funding**

The following funding sources are acknowledged:

-National Recovery and Resilience Plan (NRRP), Mission 4, Component 2, Investment 1.1, Call for tender No. 1409 published on 14.9.2022 by the Italian Ministry of University and Research (MUR), funded by the European Union – NextGenerationEU– “Metastatic potential of cancer cell aggregates in deformable matrices controlled by light” – CUP E53D23015220001. (ED, BM, SG, FG, AP).

- RSO - NTNU strategy and restructuring funds, by the Norwegian University of Science and Technology, funding a PhD fellowship under project project number 989454111, and Norges tekniske høgskoles, funding a personal “Reisestipend” travel grant to the recipient. (DU).

- Research Council of Norway, Norwegian Micro- and Nano-Fabrication Facility, NorFab, project number 295864 (ED, DU)

- National Key Research and Development Project of China (No. 2023YFF0613603).

- National Natural Science Foundation of China (No. 22202167).

- Provincial Science and Technology Plan Project: Micro and Nano Preparation and Photoelectronic Detection (No. 03014/ 1174 226063).

**Author contributions**

The authors contributed to the manuscript as follows:

D.U. and E.D. conceived the idea; D.U. fabricated the samples; B.M., and S.G. synthetized azopolymer nano-particles; E.D. performed optical actuation experiments, F.G. and A.P. performed confocal microscopy measurements and image processing; Q.M. and H.J. performed optical tweezer experiments and adhesion force measurements; E.D. performed data processing; E.D. and D.G. supervised the work; D.G., S.J., D.M. provided funding and equipment acquisition. All authors contributed to the writing and reviewing of the manuscript.

**Acknowledgements**

Not applicable